# High Througput pK$_a$ Prediction Using Semi Empirical Methods

Jan H. Jensen, Department of Chemistry, University of Copenhagen

## Significance and Impact

A large proportion of organic molecules relevant to medicine and biotechnology contain one or more ionizable groups, which means that fundamental physical and chemical properties (e.g. the charge of the molecule) depend on the pH of the surroundings via the corresponding pK$_a$ values of the molecules. As drug- and material design increasingly is being done through high throughput screens, fast - yet accurate - computational pK$_a$ prediction methods are becoming crucial to the design process. Current empirical pK$_a$ predictors are increasingly found to fail because they are being applied to parts of chemical space for which experimental parameterization data is lacking. We propose to develop a pK$_a$ predictor that, due its quantum mechanical foundation, is more generally applicable but still fast enough to be used in high throughput screening. The method has the potential to impact virtually any biotechnological design process involving organic molecules as we will demonstrate for metabolic engineering and organic battery design.

## Current State-of-the-art and Preliminary Results

There are several **empirical pK$_a$ prediction tools** (e.g. ACD pKa DB, Chemaxon, and Epik) that offer predictions in less than a second and can be used by non-experts. These methods are generally quite accurate but often fail for classes of molecules that are not found in the underlying database. Settimo *et al.*[1] have recently shown that the empirical methods are particularly prone to failure for amines (see Figure 1 for an example), which represent a large fraction of drugs currently on the market or in development. These underlying databases are not public and it is therefore difficult to anticipate when empirical methods will fail. Furthermore, the user is generally not able to augment the databases for cases where they are found to fail rendering the empirical methods essentially useless for certain molecular design projects.

pK$_a$ values can be predicted with significantly less empiricism using computational quantum mechanics (QM).[2] The accuracy of these **QM-based predictions** appear to rival that of the empirical approaches, but a direct comparison on a common set of molecules has not appeared in the literature and most QM-based pK$_a$ prediction studies have focussed on relatively small sets of simple benchmark molecules. One notable exception to the latter statement is the study by Eckert and Klamt[3] who computed the pK$_a$ by

$$\mathrm{p}K_a = c_1 \frac{\Delta G^\circ}{RT \ln(10)} + c_2 + (N_C - 1) \quad (1)$$

where $\Delta G^\circ$ denotes the change in standard free energy for the reaction

$$BH^+ + H_2O \rightleftharpoons B + H_3O^+ \quad (\text{Rxn 1})$$

and is approximated as the sum of the electronic and solvation free energy. $N_C$ - 1 is an empirical correction accounting for the observation that the the method systematically underestimated the pK$_a$ of secondary ($N_C$ = 2) and tertiary ($N_C$ = 3) amines by ca 1 and 2 pH units, respectively. Using this approach the pK$_a$ values of 58 drug-like molecules containing one or more ionizable N atoms could be reproduced with a root mean square deviation (RMSD) of 0.7. However, the method relies on conformer search at the BP/TZVP level of theory which is computationally too expensive for routine use in screening and design.

**Semiempirical QM (SQM) methods** are many orders of magnitude faster than conventional QM but their application to small molecule pK$_a$ prediction has been very limited and have focussed mainly indirect prediction using atomic charges.[4,5] The most likely reason for this is that SQM methods give significantly worse pK$_a$ predictions if used with an arbitrary reference molecule such as H$_2$O (Rxn 1). However, we[6] and others[7,8] have shown that a judicious choice of reference molecule is a very effective way of reducing the error in pK$_a$ predictions. We have very recently obtained **preliminary results** that show that this approach is the key to predict accurate pK$_a$ values using our PM6-D3H+ SQM method[9] combined with the SMD solvation method.[10] In this approach the pK$_a$ value is computed relative to some known reference value by

$$pK_a = pK_a^{ref} + \frac{\Delta G^\circ}{RT \ln(10)} + 1.5(N_C - 1) \qquad (2)$$

where $\Delta G°$ is the change in standard free energy of the reaction

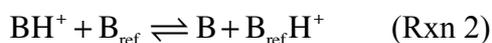

$$BH^+ + B_{ref} \rightleftharpoons B + B_{ref}H^+ \qquad (Rxn\ 2)$$

This PM6-D3H+/SMD approach yields a mean absolute deviation (MAD) of 1.2 and 0.6 for carboxylic acids and primary amines (using ethanoic acid and ethylamine as references), which compare well with the values of 0.7 and 0.3 computed by Satre *et al.*[11] using the thousand-times more expensive M05-2X/6-311++G(d,p)/SMD method. The amine test set used by Satre *et al.* contained only one example each of a secondary and tertiary amine, so we created additional test sets for each amine-type and found that PM6-D3H+/SMD pK$_a$ predictions, using di- and triethylamine as respective references, resulted in pK$_a$ values that are systematically underestimated by about 1.5 and 3 pK$_a$ units, respectively. But by introducing the correction suggested by Eckert and Klamt the PM6-D3H+/SMD MADs are 0.7 for both secondary and tertiary amines.

Encouraged by these results we applied the PM6-D3H+/SMD approach to compound **1**, for which Settimo *et al.*[1] have shown that the popular Chemaxon method predicts a pK$_a$ value of 9.1, which is significantly higher than the experimental value of 4.2, i.e. Chemaxon predicts that **1** is charged as physiological pH when, in fact, it is neutral. PM6-D3H+/SMD predicts a pK$_a$ value of 0.1 (i.e. neutral as physiological pH) using triethylamine as a reference. Furthermore, unlike for Chemaxon and related methods *the accuracy can be improved* by using reference molecules that are more similar to the target (Figure 1). The computational cost of computing the free energy of a single conformation of **1** is ca 5 minutes on a single Intel Xeon 2.67GHz X5550 core processor, making high throughput pK$_a$ prediction possible.

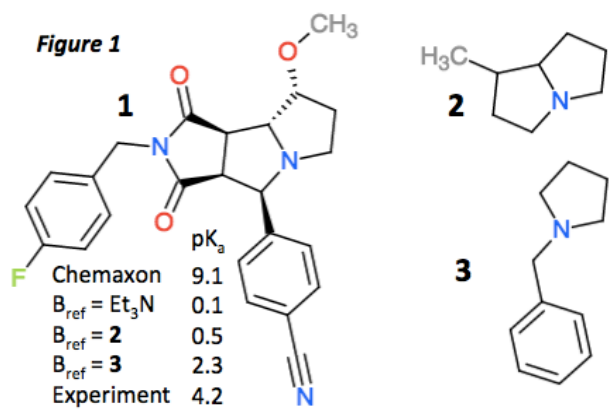

## Milestones

**Milestone 1:** *A robust, automated SQM-based pK$_a$ predictor by late-2017*

**Milestone 2:** *A more accurate SQM-based pK$_a$ predictor by mid-2018*

**Milestone 3:** *Application to high throughput screen of new organic batteries by mid-2019.*

## Research Plan and Methodology

**Work Package 1: Development of a robust, automated PM6-D3H+/SMD pK$_a$ predictor**

**WP1a: Implementation of PM6 and PM7 in GAMESS**. PM6-D3H+ and the interface to SMD is implemented in the GAMESS program, but only for elements up to F. The heavier elements require new integral code, which MOPAC-developer James Stewart has supplied and which will be implemented in GAMESS and adapted to the SQM/PCM interface developed by us.[12] The PM7 parameter set will also be implemented. **WP1b: Reoptimization of SMD parameters for PM6 and PM7**. The reoptimization will be done in collaboration with the SMD developers (see letter from Cramer and Truhlar) and should increase the accuracy of the pK$_a$ predictions made, so far, with SMD parameters optimized for conventional QM methods. We will test both PM6 and PM7 and anticipate that a common SMD parameter set can be found that can also be used with most any dispersion and hydrogen bonding correction. **WP1c: Automation of the pK$_a$ prediction workflow for high throughput studies and use by non-experts**. All steps in the workflow will be automated including initial coordinate and conformer generation, error detection, coupled titration,[13] and identification of a suitable pK$_a$ reference using similarity searches.[14] **WP1d: Extensive benchmarking**. With the automation in place we will carefully benchmark both PM6- and PM7-based methods against experimental data for most of the common functional groups including groups like sulfonic and phosphonic acids and sulfides which are not well studied by QM-based pK$_a$ predictors. Rules for determining the best choice of reference (B$_{ref}$, Rxn 2) will be developed as part of this work. **WP1e: Web interface.** We have found that an easy-to-use-web interface is a powerful way to introduce potential users to a new computational methodology and increase the user base. For example, in 2005 we launched the first web interface for protein $pK_a$ prediction based on our PROPKA approach and, largely as a result of this, the original PROPKA paper[15] has now been cited close to 900 times. Accordingly, we will create a web-interface for the PM6-D3H+/SMD $pK_a$ predictor. This will be relatively easy as the we have already developed a web interface for automated GAMESS calculations for teaching at molcalc.org.[16]

### Work Package 2: Increasing accuracy using machine learning

Recent work by von Lilienfeld[17] has shown that the performance of SQM methods can be improved significantly by machine learning. For example, the G4MP2 atomization enthalpies of 6095 constitutional isomers of $C_7H_{10}O_2$ can be reproduced with a MAD of 2.4 kcal/mol using PM7, compared to an MAD of 6.4 kcal/mol for uncorrected PM7, using a the Δ-machine learning approach and a training set of 2000 molecules. It is therefore likely that the accuracy SQM-based $pK_a$ predictions can be improved significantly by training against corresponding QM results. **WP2a**: We will benchmark several QM-based $pK_a$-prediction schemes (e.g. Ref 3 and 11) and select the best compromise between accuracy and speed. **WP2b:** Then we will generate a training set of 2000-5000 molecules covering diverse set of ionizable groups in a diverse set of chemical environments, compute QM-based $pK_a$ values, and train SQM-based $pK_a$ predictions, using Δ-machine learning approach and in collaboration with von Lilienfeld (see letter), against half of the set and use the other half for validation. Following Rupp *et al.*[18] we will impose locality to ensure transferability. **WP2c**: The resulting SQM-based method is then benchmarked against experimental data as before (WP1d). Because we train against QM-calculations the machine learning approach will have an abundance of high-quality relevant data with little or no noise - since the output of the calculation is the precise value that the machine learning method is trying to predict.

### Work Package 3: Application to two high throughput studies

**WP3a: Metabolic engineering**, optimizing genetic and regulatory chemical networks within cells to increase the cells' production of a certain substance, is key to modern biotechnological production. However, in order to model and rationally engineer metabolic pathways, standard Gibbs reaction energies are needed for the reactions in the chemical network and many of these values are not known experimentally. Aspuru-Guzik and co-workers[19] have initiated an ambitious project to fill in the missing thermodynamic data with data computed using high throughput QM calculations. However, many of the compounds in the study contain ionizable groups whose charge state greatly influences the computed thermodynamic data. We will therefore, in collaboration with Aspuru-Guzik (see letter), interface our SQM-$pK_a$ prediction approach with these high throughput calculations in order to predict the correct protonation state at a relevant pH values and, conversely, to predict the effect of pH on the metabolic pathways.

**WP3b: Computational design of organic molecules for an aqueous flow battery.** Replacing expensive redox-active metals in aqueous flow batteries with abundant, carbon-based molecules, such as quinones, can dramatically lower the cost of electricity storage, because they can potentially be sourced from biological sources. Identifying suitable organic molecules is therefore a high priority and significant challenge given the vast size of chemical space to be explored. For example, Aspuru-Guzik has lead efforts to computationally screen over 3.5 million molecules using a specialized distributed computing framework to compute the 30,000 CPU years expended.[20] As with WP3a, many of the compounds in the study contain ionizable groups whose charge state greatly influences the computed properties. Thus, in collaboration with Aspuru-Guzik, we will interface our SQM-p$K_a$ prediction approach with these high throughput calculations in order to predict the correct protonation state at a relevant pH values and, conversely, to predict the effect of pH on the redox properties.

**Work Plan**

| Month | 6 | 12 | 18 | M | 24 | M | 30 | 36 | M |
|---|---|---|---|---|---|---|---|---|---|
| WP1a |  |  |  | I |  | I |  |  | I |
| WP1b |  |  |  | L |  | L |  |  | L |
| WP1c |  |  |  | E |  | E |  |  | E |
| WP1d |  |  |  | S |  | S |  |  | S |
| WP1e |  |  |  | T |  | T |  |  | T |
| WP2a |  |  |  | O |  | O |  |  | O |
| WP2b |  |  |  | N |  | N |  |  | N |
| WP2c |  |  |  | E |  | E |  |  | E |
| WP3a |  |  |  |  |  |  |  |  |  |
| WP3b |  |  |  | 1 |  | 2 |  |  | 3 |

**Facilities.** We have access to 1500 cores as part of the University of Copenhagen branch of the Danish Supercomputer Center. The DCSC provides a full-time TAP for the day-to-day running of the DCSC/KU center. Jensen and co-worker recently performed a study[21] involving 2.4 million DFT calculations using this facility, which is therefore also adequate for the current proposed work.

**Publication and Dissemination.** All theoretical developments and applications will be published in peer-reviewed journals. We will publish in open access journals or journals with an open access option, to allow access to as many people as possible. All software will be made freely available to the scientific community as a user-friendly software package and as a web interface (WP1e).

**Ethical Aspects.** The proposed work will not involve the use of laboratory animals.